# 1 METHODOLOGY

The research methodology is essentially the strategy used to achieve the objectives of the research, by using one or more research methods. [1-3] categorised three, broadly generic, research strategies:

- Experiments, through the manipulation of variables;
- Surveys and interviews, through the collection of information from individuals or groups;
- Case study, based on intensive knowledge of about one case of a number of related cases;

These will be further detailed in the subsequent sub-sections. Prior to these, [4] identifies two other major dimensions that need to be defined in the methodology section of a doctoral dissertation, namely the research philosophy and the reasoning of the research.

## 1.1 RESEARCH PHILOSOPHY

According to [5], a research paradigm is "the set of common beliefs and agreements shared between scientists about how problems should be understood and addressed" and it involves the following philosophical concepts [6]:

- Ontology – branch of metaphysics that questions what is the nature of existence.
- Epistemology – branch of philosophy that deals with how we go about knowing something and what constitutes valid knowledge and how can we obtain such knowledge.

These concepts create a holistic view of the knowledge we seek and the methodological strategies at our disposable to uncover such knowledge. Table I (adapted from [7-9]) summarizes the three main research paradigms in terms of its ontological and epistemological perspectives and the used methods:

*Table I – Research paradigms*

| Paradigm | Ontology | Epistemology | Method |
|---|---|---|---|
| Positivism | There is a single reality or truth | Reality can be measured through proper tools | Quantitative (e.g. statistical analysis) |
| Constructivism | There is no single reality of truth as reality is created by individuals in groups | Reality needs to be interpreted. We need to find the underlying meaning of a scenario | Qualitative (e.g. interviews, case study, etc.) |
| Pragmatism | Reality is constantly renegotiated in light of new situations | The best method is the one that solves problems | Mixed (combination of the above) |

## 1.2 THE REASONING OF THE RESEARCH

For many social scientists, social change is fuelled by technological change theory, as economic and technological materialistic factors are what shape humanity and social systems are determined by technological systems. Not only technology gives humanity the ability to create and utilise energy, it also holds a crucial social feature: information – its amounts and uses – as it is well established that the more information a society holds, the more evolved and advanced it is and this translates into advancements in the economic system and political system, distribution of goods, social inequality and other spheres of social life.

Many features of the social sciences, especially in philosophy of science [10, 11], take a particular place on this thesis, even if in an indirectly manner:

- According to Darwinian evolution and technological change, the individuals that compose a society must evolve with the society or they will be left behind and be abandoned brutally by that society.

- Holism tells us that the hole is more than the sum of its parts. This is particularly true when comparing a society who lacks access to information with a highly technologically one: every individual takes a benefit from information access, but the sharing of information (and the consequently disproved information) provides the whole society with more – and with better quality – information that the sum of individual information.
- Hermeneutics talk about the use of rules. This is also applied in a technologically advanced society: the access to information is guided by certain rules, which may or not be optional, including rules imposed by telecommunications regulatory authorities.
- Rational choice theory defends that an agent obtains a certain amount of utility (or satisfaction) from any amount of commodity, subject to the limitations on available resources and information, by creating a portfolio of commodities that maximize his utility. Telecommunications systems also have a certain utility value and a perceived value for which agents will have to balance against other alternative commodity choices.
- Critical theory provides guidance about the way the world ought to be. This is also true in a technologically advanced society: it is a fact that information access helps highly advanced technological societies be shaped in a certain direction which translates in better educational and healthcare services.

Several other examples could be applied, but the main point is that, more than a natural science, access to information technology in a society has its roots on social scientific facts and it is guided by the beliefs and desires of society. And this is what drives the need for telecommunications and computer systems.

## 1.3   TYPES OF RESEARCH

One can identify two major types of research methodologies: the deductive method) and the inductive. The former, predominately used in natural sciences, starts by analysing the current state-of-the-art (this provides the much needed context for the research), and then proceed to identify a problem for which the current state-of-the-art does not provide an adequate answer (this could be for example a difference between theory and evidence; some founded contradictions or a new context for the application of previous findings) [12]. The later – inductive research – is predominantly used in the social sciences to gain explanation or understanding through empirical observations. In this particular type of research, the research abstains from formulating a hypothesis a priori, unlike in the deductive research, where the researcher basis his research and data collection in the current body of knowledge [13]. Table II [14] compares both methodologies:

*Table II – Comparison of deductive and inductive research*

| Type of research | **Deductive** | **Inductive** |
|---|---|---|
| Explanation | Via analysis of causal relationships | Of subjective meaning |
| Data | Quantitative | Qualitative |
| Methods | Use of physical or statistical controls, to allow hypothesis testing, uses closed-ended questions, pre-determined approaches, numeric data | Real world research, attempts to minimise reactivity among the research subjects, uses open-ended questions, emerging approaches, text and/or image data |
| Practices | Tests or verifies theories or explanations<br><br>Identifies variables of interest<br><br>Relates variables in questions or hypotheses<br><br>Uses standards of reliability and validity<br><br>Observes and then measures information numerically<br><br>Uses unbiased approaches<br><br>Employs statistical procedures | Positions researcher within the context<br><br>Collects participant-generate meanings<br><br>Focuses on a single concept or phenomenon<br><br>Brings personal values into the study<br><br>Studies the context or setting of participants<br><br>Validates the accuracy of findings<br><br>Interprets the data<br><br>Creates an agenda for change or reform<br><br>Involves researcher in collaborating with participants |

One can also divide the types of research into two small groups [15]:

- Basic research: refers to the study of a phenomenon in order to better understand it. Implies the formulation of theories with the aim of advancing knowledge, while the practical application of the results is relegated to a second plan.
- Applied research: refers to the acquisition of knowledge, based on an already build-up theoretical foundation, to solve a practical problem. It is generally based on a commitment to promote public welfare through a cost-effective reduction of a social issue, based in a real-world setting.

It is important to note that a mix usage of quantitative and qualitative data can also be used in some cases. Besides the straight-forward research types explained previously, one can describe the type of research in other terms, specifically:

- Exploratory research: aims at the formulation of problems and hypotheses, rather than test them, using qualitative data.
- Explanatory research: uses quantitative data in statistical tests to explain a phenomenon and predict future occurrences, through research hypotheses that specify the nature and relationships between the variables under study. This type of research does not intend to provide a conclusion on the topic, but instead it helps the researcher to understand the problem more efficiently.
- Descriptive research: uses data collected from samples (can be either qualitative or quantitative or both) to describe and characterize persons, goods or scenarios, through statistical sampling. Four methods can be used in descriptive research [16]:
    - Correlational: the researcher observes statistical correlational between variables in order to establish whether they are related;
    - Development design: the researcher learns how variables change over time;
    - Observational: the researcher observes and documents human behaviour;
    - Survey: the researchers collect sampling data from respondents;

## 1.4 TYPES OF DATA

Data can simply be divided into two major categories: quantitative data and qualitative data. The former is typically factual data and it is used to study relationships between facts and provide conclusions in the light of the current state-of-the-art [17] and are assumed to be repeatable, while the later – qualitative data – is used to evaluate and theorize problems and approaches [18], focusing in the qualities of the investigated phenomena rather than its numeric measurement. Then, data can be of two types [19]: primary – if collected directly by the researcher (e.g. through surveys) – or secondary – if previously collected by someone else (e.g. historical statistics).

## 1.5 DATA COLLECTION

There are three main ways of collecting data:

1. Collect data from observation, either from nature or from conducted experiments – popular methodologies in exact or natural sciences;
2. Obtain data from individuals or communities through interviews, surveys or other related methodologies – popular methodologies in the humanities and social sciences;
3. Use historical data – popular methodologies in all sciences;

For the obtained data to be considered 'good', it needs to simultaneously be obtained from appropriated sources and in a very large scale, as we could see in [20], the consulted audience needs to be broad, credible and of diverse roles.

A concern regarding the previous point 2 is that in this way of collecting data, this data is purely subjective and can easily be manipulated by, e.g., removing the outliers from the data, in order to obtain the desired results (in fact, the reproducibility of results is a major disadvantage of this method). Not to mention the fact that researchers can make their own interpretation of the respondents' words, which can ultimately affect the results and the overall quality of the research, as stated by [21]. Another concern is that, since the data is subjective, if someone else replicates the surveys, the results will naturally be different. Not to mention other concerns that would arise if this way of obtaining data was chosen, such as the disclosure of sensitive data from respondents and the identify protection concerns as stated by [22].

## 1.6 RESEARCH DESIGN

The research design of a quantitative project as a step-wise approach is with the following sequences:

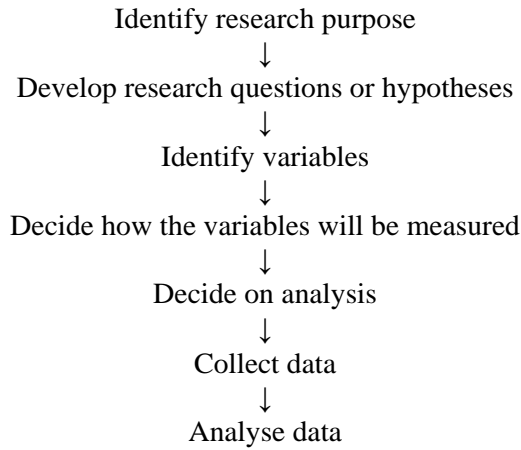

## 1.7 ETHICAL CONSIDERATIONS

Just like many professions have to deal with deontological concerns on a daily basis, so does a researcher, who has to comply with ethical concerns. The most obvious ones are to make sure the work is not plagiarized, that the work is original (otherwise it would not be called research), that the works of others are acknowledged and that these 'others' will not see their unpublished work or patents disclaimed without permission. These seem rather obvious and it is expected that every researcher complies with them. Another consideration that the researcher must have, is that the results are documented in a manner that another research can re-create the results using the methodology of the dissertation, in line with the guidelines provided by [23].